\documentclass[preprint,12pt,numbers]{elsarticle}

\usepackage{amssymb}
\DeclareMathSymbol{\shortminus}{\mathbin}{AMSa}{"39} 

\usepackage{amsmath}
\usepackage{amsthm}
\allowdisplaybreaks
\newtheorem{lem}{Lemma}
\theoremstyle{definition}
\newtheorem{example}{Example}
\newtheorem{remark}{Remark}

\usepackage{lineno}
\usepackage{etoolbox} 

\newcommand*\linenomathpatch[1]{%
  \cspreto{#1}{\linenomath}%
  \cspreto{#1*}{\linenomath}%
  \csappto{end#1}{\endlinenomath}%
  \csappto{end#1*}{\endlinenomath}%
}

\linenomathpatch{equation}
\linenomathpatch{gather}
\linenomathpatch{multline}
\linenomathpatch{align}
\linenomathpatch{alignat}
\linenomathpatch{flalign}

\usepackage{todonotes}
\usepackage{mathtools}
\usepackage{booktabs}
\usepackage{diagbox}

\usepackage{xcolor}
\definecolor{cellgray1}{gray}{0.9}
\definecolor{cellgray2}{gray}{0.5}
\makeatletter
\protected\def\ccell#1#{%
  \kern-\fboxsep
  \@ccell{#1}%
}
\def\@ccell#1#2#3{%
  \colorbox#1{#2}{#3}%
  \kern-\fboxsep
}
\makeatother
\usepackage{csquotes}
\usepackage{acronym}
\usepackage{dsfont}

\usepackage{tikz}
\usetikzlibrary{calc}
\usepackage{csvsimple}
\usepackage{float}
\usepackage{makecell} 
\usepackage{orcidlink}
\usepackage{fontawesome}
\newcommand{\github}[1]{%
   \href{#1}{\faGithub}%
}
\usepackage{multirow}
\usepackage[T1]{fontenc}

\usepackage{changes}
\newcommand{\stkout}[1]{\ifmmode\text{\sout{\ensuremath{#1}}}\else\sout{#1}\fi}
\setdeletedmarkup{\stkout{#1}}

\newcommand{\myitem}[2]{{\makebox[0pt][l]{#1}\phantom{#2}}}

\makeatletter
\newenvironment{varsubequations}[1]
 {%
  \addtocounter{equation}{-1}%
  \begin{subequations}
  \renewcommand{\theparentequation}{#1}%
  \def\@currentlabel{#1}%
 }
 {%
  \end{subequations}\ignorespacesafterend
 }
\makeatother

\acrodef{mip}[MIP]{mixed-integer linear programming}
\acrodef{sm}[RCPSP]{resource-constrained project scheduling problem}
\acrodef{mm}[MRCPSP]{multi-mode resource-constrained project scheduling problem}
\acrodef{mn}[MRCPSP/N]{}
\acrodef{mr}[MRCPSP/R]{}
\acrodef{ma}[MAP]{mode assignment problem}
\acrodef{se}[SEE]{start-end event}
\acrodef{sea}[SEE-A]{}
\acrodef{rse}[RSEE]{revised start-end-event}
\acrodef{ie}[IE]{interval event}
\acrodef{oo}[OOE]{on-off event}
\acrodef{ooa}[OOE-A]{}
\acrodef{ct}[CT model]{}
\acrodef{dt}[DT model]{}
\acrodef{ctab}[MCTAB-EXT]{MCTAB extended}
\acrodef{fct}[FCT]{}
\acrodef{fctw}[FCT-W]{weak linear flow-based model}
\acrodef{fcts}[FCT-S]{strong linear flow-based model}
\acrodef{fcto}[FCT-OZON]{}
\acrodef{psplib}[PSPLIB]{project scheduling problem library}
\acrodef{aon}[AON]{activity-on-node}

\newcommand{\bigO}{\mathcal{O}}

\newcommand{\ance}{\ensuremath{\mathrm{\bar A}}}
\newcommand{\desc}{\ensuremath{\mathrm{\bar D}}}

\makeatletter
\newcommand{\smallplus}{\mathbin{\mathpalette\make@small +}}
\newcommand{\smallminus}{\mathbin{\mathpalette\make@small -}}

\newcommand{\make@small}[2]{%
  \vcenter{\hbox{%
    $\m@th\ifx#1\displaystyle\scriptscriptstyle\else\ifx#1\textstyle\scriptscriptstyle
     \else\scriptscriptstyle\fi\fi#2$%
  }}%
}
\makeatother

\def\set#1{\mathcal{#1}}
\def\sSetminus#1#2{#1^{\smallminus#2}}
\def\sSetplus#1#2{#1^{\smallplus#2}}
\def\sA{\set{A}}
\def\sR{\set{R}}
\def\sN{\set{N}}
\def\sP{\set{P}}

\def\sM{\set{M}}
\def\sMi{\sM_i}
\def\sV{V}
\def\sE{\set{E}}
\def\sH{\set{H}_G}

\def\sEE{\sE^2}
\def\sPrec{\sP}
\def\sPrecCore{\bar{\sP}}
\def\Gbar{\bar{G}}
\def\sAncestors#1{\ance(#1)}
\def\sDescendants#1{\desc(#1)}

\def\cardA{{A}}
\def\terminalA{{\bar{\cardA}}}

\def\cardR{{R}}
\def\cardN{{N}}

\def\makespan{C_\text{max}}
\def\bigM{M}
\def\bigMf{\bigM^f_{ijk}}
\def\bigMs{\bigM^s_{ij}}
\def\sS{\set{S}}
\def\varx{\tilde x}
\def\vary{\tilde y}

\begin{document}

\begin{frontmatter}



\title{Continuous-Time Formulations for Multi-Mode Project Scheduling}


\author[inst1]{David Sayah\corref{cor1}}
\cortext[cor1]{Corresponding author at: Haid-und-Neu-Str.10-14, 76131, Karlsruhe, Germany~\orcidlink{0000-0002-8977-9414}~\github{https://github.com/ddotsdot/}}
\ead{sayah@fzi.de}

\affiliation[inst1]{organization={FZI Research Center for Information Technology},
            postcode={D-76131},
						city={Karlsruhe},
            country={Germany}}



\begin{abstract}
This paper reviews compact continuous-time formulations for the multi-mode resource-constrained project scheduling problem. Specifically, we first point out a serious flaw in an existing start-end-event-based formulation owing to inconsistent mode choices. We propose two options to formulate the missing constraints and consider an equivalent reformulation with sparser constraint matrix. Second, we formulate an aggregate variant of an existing model that relies on on-off-events, and we clarify the role of mode consistency issues in such models. Third, we suggest two variants of an existing network flow formulation. We enhance our models by adapting several techniques that have been used previously, e.g., in cases with only a single mode.
A large set of benchmark instances from the literature provides the basis for an up-to-date and fair computational study with an out-of-the-box solver package. We compare our models against two models from the literature. Our experiments assert confidently that network flow formulations prevail in the test bed, and they provide a hint on why event-based models become less competitive in multi-mode settings.
\end{abstract}



\end{frontmatter}

\section{Introduction}
\label{sec:intro}
Project planners face the \ac{sm} when they want to schedule a given set of activities, i.e., determine a starting time for each activity subject to given precedences between some activities and resources with limited capacity usage rates. Each activity is associated with a given process duration and consumption rates for the required resources. The objective most commonly pursued in this context is to minimize the overall project completion time, called \emph{makespan}.

This paper focuses on the multi-mode extension of \ac{sm} (\acused{mm}\ac{mm}), which is known to be $\mathcal{NP}$-hard \citep{Blazewicz1983}. It comprises deciding not only when but also how the activities should be performed. An activity can be accomplished in one of multiple predefined ways, or, \emph{operational modes}, each describing a distinct time-resource combination. \ac{mm} therefore incorporates various time-resource and/or resource-resource tradeoffs \citep{Sprecher1998}. 

In addition to the so-called renewable resource type considered in \acp{sm}, \acp{mm} account for \emph{non-renewable resources} that limit the total capacity usage of each resource over the entire project lifetime \citep{Talbot1982}. As noted by \cite{Kolisch1995a}, broadening the scope of project scheduling in this way entails solving a \acfi{ma} on top of an \ac{sm}. Instances of \ac{mm} that entirely ignore non-renewable resources are referred to as \acused{mr}\ac{mr} \citep[see also][]{Peteghem2014} and those with only non-renewable resources as \acused{mn}\ac{mn}. A detailed overview of the existing body of literature on \acp{mm} is presented in the comprehensive and newly updated survey of \cite{Hartmann2022}.

Our paper specifically addresses continuous-time (CT) formulations of \ac{mm}. This modelling approach formulates the starting times in continuous space. On the contrary, discrete-time (DT) formulations assume a discrete planning horizon and they typically involve time-indexed binary variables for each activity. \acused{dt}\acp{dt} have been dealt with at length in the scheduling literature (for instance, the classical models developed by \cite{Pritsker1969,Christofides1987,Talbot1982} or the models recently reviewed in the work of \cite{Artigues2017}).
\acused{ct}\acp{ct} for \ac{sm} have been discussed in the literature for quite some time \citep{Artigues2003,Kone2011}. As an example,  \cite{Artigues2003} cast  \ac{sm} as a network flow optimization problem and formulate it as a \ac{mip} problem. \cite{Kone2011} develop two \ac{mip} formulations based on the notion of \acfi{se} and \acfi{oo}.

All the above cited \acp{ct} are well-known for the fact that their linear programming (LP) relaxations are weak compared to those of \acp{dt} \citep[e.g.][]{Demassey2005,Kone2011,Artigues2013,Tesch2020}. This is largely due to the presence of \enquote{bigM}-constraints. Nevertheless, a striking advantage of \acp{ct}, apart from their capability to cope with non-integer activity durations, is the fact that their model size is insensitive to variations in time-related input data.
In contrast, the number of variables and constraints in the \ac{dt} of \cite{Pritsker1969}, for example,  increases pseudo-polynomially in the number of activities and time periods when activity durations (hence the planning horizon) increase.
In the computational study conducted by \cite{Kone2011}, the authors compare the relative performance of both DT and \acp{ct} using benchmark instances from the literature whose activity durations were scaled up.
Interestingly, they report that in these instances \acp{ct} are superior to \acp{dt} in terms of integer solving performance despite poor LP relaxations.

Some attempts to model the \ac{mm} in continuous time have been made to date. On the one hand, the \ac{mip} models of \cite{Kyriakidis2012} and \cite{Gnaegi2019} are to be noticed here. The model of \cite{Kyriakidis2012} assumes a so-called resource-task network (RTN) representation of the problem and  the planning horizon is divided into subintervals of variable lengths. The \emph{assignment-based} \acp{ct} of \cite{Gnaegi2019} assume discrete resource capacities and assign each activity to one or more capacity units without overlap, an idea similar to the one presented earlier by \cite{Correia2012} for the \ac{sm} variant with so-called flexible resources. The computational study in \cite{Gnaegi2019} shows that, provided the range of durations is relatively large, the assignment-based \acp{ct} outperform the \ac{dt} of \cite{Talbot1982} and, in particular, the \acp{ct} of \cite{Kyriakidis2012}.

On the other hand, a few event-based models haven been proposed. \cite{Zapata2008} formulate a mathematical model for multi-project \acp{mm} using \acp{se} (in fact, the \ac{se} model of \cite{Kone2011} is an improved version for \acp{sm} with less variables). \cite{Chakrabortty2014} extended \citeauthor{Kone2011}'s \ac{se}-based and \ac{oo}-based models to handle multiple operational modes. The former incorporates non-renewable resources, while the latter ignores them. 
Recently, \cite{Ozturk2020} considered \ac{mr} and presented an \ac{se}-based model, which in fact resembles that of \cite{Chakrabortty2014}, and they propose a new flow-based model.
Table~\ref{tab:refs} summarizes the references.

\begin{table}[htbp]
    \centering
    \small
    \begin{tabular}{lccc}
    \toprule
         \diagbox{model}{problem} & \ac{mr} & \ac{mm} \\
         \midrule
         \ac{se}-based & \ccell{cellgray1}{\,\,~\cite{Ozturk2020}~} & \ccell{cellgray1}{\cite{Chakrabortty2014}}\\
         \ac{oo}-based & \ccell{cellgray1}{\cite{Chakrabortty2014}} & \ccell{cellgray1}{\phantom{\cite{Chakrabortty2014}}} \\
         flow-based & \ccell{cellgray1}{\,\,~\cite{Ozturk2020}~} & \ccell{cellgray1}{\phantom{\cite{Chakrabortty2014}}} \\
         assignment-based &  & \cite{Gnaegi2019}  \\
         resource-task network &  & \cite{Kyriakidis2012}  \\
    \bottomrule
    \end{tabular}
    \caption{Overview of references related to \acp{ct} for \ac{mm} and \ac{mr}; highlighted cells  ( \ccell{cellgray1}{\phantom{--}} ) indicate the contribution of this paper}
    \label{tab:refs}
\end{table}

A remarkable advantage of \ac{se}- and flow-based models is that model size is insensitive to changes in capacity-related problem parameters, whereas assignment-based models grow considerably fast in size as the range of capacity requirements increases. Specifically, assignment-based models employ capacity-indexed binary variables for each activity and each renewable resource. The number of variables and constraints therefore increases pseudo-polynomially in the number of activities, the number of renewable resources, and the number of capacity units.  

Having said that, it seems virtually impossible to assess the relative performance of existing \acp{ct} for \ac{mm} without computational experiments taking this crucial difference into account. 
Unfortunately, current literature does not provide such a dedicated numerical comparison. Even worse, the \ac{se}-based models as presented in \cite{Chakrabortty2014} for \ac{mm} and in \cite{Ozturk2020} are incomplete and can thus produce infeasible schedules.

Our contribution is as follows. We begin with a proper counterexample pointing out that the \ac{se}-based models suggested in \citep{Chakrabortty2014,Ozturk2020} are missing out an important set of constraints needed to guarantee a consistent mode choice for each activity. We formulate the missing constraints and discuss possible merits of mode consistency constraints in \ac{oo}-based models. We also present aggregate versions of both the \ac{se} and \ac{oo} model.
Second, we derive an new variant of the \ac{se} model with sparser constraint matrix.
Third, we provide a weaker and stronger variant of the network flow model presented in \cite{Ozturk2020}.
We enhance our models by adapting previously known techniques which aim to (i) incorporate additional inequalities, e.g., time window constraints, enforced sequential processing for incompatible activities, and (ii) to fix variables.
Finally, we set up a computational study using more than 3200 benchmark problems, partly taken from the \ac{psplib} \citep{Kolisch1997} and the MMLIB \citep{Peteghem2014} and partly derived by scaling up the demands and capacities of renewable resources. Our aim is to fairly evaluate our proposed event-based and flow-based models in comparison with the assignment-based approach. Additionally, we set up a second group of experiments with another 1594 instances of \ac{mn} that we derived from the \ac{psplib}. These experiments aim to examine the impact of an increasing number of modes on the computational tractability of the models.

The remainder of this paper is organized as follows. In Section~\ref{sec:prob}, we briefly introduce a formal problem description. Section~\ref{sec:models} presents our event-based and flow-based formulations of \ac{mm}, while the simple enhancement techniques are discussed in Section~\ref{sec:enhancements}. We summarize the findings based on our numerical results in Section~\ref{sec:experiments} and conclude this paper with Section~\ref{sec:conclusion}.

\section{Problem statement}
\label{sec:prob}
We start with general parameters that are given in the problem at hand.
Let $\sA=\{1,\dots,\cardA\}$ denote a set of activities, indexed $i$ and $j$. Each activity $i$ is associated with a set of operational modes $\sMi\subset\{1,2,\dots\}$, indexed $m$, in which the activity can be executed. 

Moreover, we are given the amount of time~$p_{im}$ it takes to process activity $i\in\sA$ in mode $m\in\sMi$. Let $\sPrecCore$ be a given set of precedence relationships between non-dummy activities. That means if activity $i$ must be completed before (not necessarily immediately) activity $j$ can start, then the ordered pair $(i,j)$ is in $\sPrecCore$.

Let~$\sR=\{1,\dots,\cardR\}$ be a given set of renewable resources, indexed by~$k$. Each resource provides $B_k$ units of capacity per unit of time. Likewise, a set~$\sN=\{1,\dots,\cardN\}$ of non-renewable resources is given where each non-renewable resource provides $W_k$ units of capacity over the entire project lifetime. Processing an activity $i$ in mode $m$ requires $b_{imk}$ capacity units per time unit from renewable resource $k\in\sR$ and $w_{imk}$ capacity units per project of non-renewable resource $k\in\sN$.
For convenience, we assume~$b_{imk}\leq B_k$ for all~$i\in\sA,m\in\sMi,k\in\sR$ and~$w_{imk}\leq W_k$ for all~$i\in\sA,m\in\sMi,k\in\sN$ (i.e., exclusion of infeasible activity-mode combinations).

The nonpreemptive \acf{mm} is about determining, for each activity~$i\in\sA$, a starting time $S_i$ and one operational mode~$m(i)$ out of~$\sMi$ such that
\begin{itemize}
    \item the total mode-dependent capacity requirements meet the available capacity for each resource type at any time,
    \item the given precedence relationships between activities are respected, i.e., $S_i + p_{i,m(i)} \leq S_j$ for each~$(i,j)\in\sPrecCore$,
    \item activities are processed without interruption,
    \item the makespan~$\makespan$ defined as the maximum completion time $C_i=S_i+p_{i,m(i)}$ of any activity $i\in\sA$ is minimized.
\end{itemize}
When we say \enquote{(in)consistent mode choice}, we refer to the above requirement that each activity $i\in\sA$ must be assigned exactly one mode~$m(i)$. Moreover, we refer to instances of \ac{mm} with only non-renewable resources as \acused{mn}\ac{mn}. Note that this special case still contains an \ac{ma} as a feasibility subproblem. Its $\mathcal{NP}$-completeness was established in \cite{Kolisch1995a}, given that there are at least two non-renewable resources and at least two modes for each activity. 
Hence, \ac{mn} remains $\mathcal{NP}$-hard.

\section{Continuous-time formulations}
\label{sec:models}
\subsection{Start-end-event formulation}
\label{sec:ct-see}

A \acf{se} as defined in \citep{Kone2011} associates the start and completion times of one or more activities with an event. This concept exploits the fact that for \ac{sm} there exists always an optimal solution, in which the start time of an activity is either equal to the completion time of another activity or 0 (\emph{left-shifted schedule}). Thus, at most $\cardA+1$ \acp{se} need to be considered and the corresponding index set is given by~$\sE=\{0,1,\dots,\cardA\}$.
The notation
\[
\sEE :=\left\{(e,f)\in\sE\times\sE: e<f\right\}
\]
refers to the set of pairs of consecutive events. The shorthand notation~$\sSetminus{\set{S}}{e}$ ($\sSetplus{\set{S}}{e}$)  comes in handy when we intend to denote the removal (addition) of one element from (to) a general set~$\set{S}$.

In the \ac{se}-based models of \cite{Chakrabortty2014} and \cite{Ozturk2020}, two types of binary decision variables are defined for each $i\in\sA,m\in\sMi,e\in\sE$: 
$x_{ime}=1$ and $y_{ime}=1$ indicate if and only if processing activity $i$ in mode~$m$ starts, respectively, finishes at event $e$.
A continuous variable $s_e$ is used to measure the point in time when event $e\in\sE$ occurs. It is generally assumed that $s_0\leq s_1\leq\cdots\leq s_\cardA$ holds such that~$s_0$ is the start time of a first activity and~$s_\cardA$ the end time of a last activity (i.e., the makespan). A continuous auxiliary variable $r_{ek}$ keeps track of the amount of capacity of resource $k\in\sR$ that must be available at the time of event $e\in\sE$. We consider the following \ac{ct} formulation:
\begin{subequations}
\label{ct-see}
\begin{align}
    z^{\acs{se}} ={}& \min s_\cardA \label{ct-see-obj}\\
    \text{s.t. }&%
    s_0 = 0 \label{ct-see-first-event-start}\\
    & s_e \leq s_{e+1} \quad e\in\sSetminus{\sE}{\cardA} \label{ct-see-event-order}\\
    & s_f \geq s_e + p_{im}(x_{ime} + y_{imf} - 1) \quad i\in\sA,m\in\sMi,(e,f)\in\sEE \label{ct-see-event-time-def}\\
    & \sum_{m\in\sMi} \sum_{e\in\sSetminus{\sE}{\cardA}} x_{ime} = 1 \qquad i\in\sA \label{ct-see-start-event}\\
    & \sum_{m\in\sMi} \sum_{e\in\sSetminus{\sE}{0}} y_{ime} = 1 \qquad i\in\sA \label{ct-see-end-event}\\
    & \sum_{m\in\sMi}\sum_{e'=1}^{e} y_{ime'} + \sum_{m\in\sMi}\sum_{e'=e}^{\cardA-1} x_{ime'} \leq 1 \quad i\in\sA,e\in\sE\setminus\{0,\cardA\} \label{ct-see-precedence2}\\
    & \sum_{m\in\sMi} \sum_{e'=e}^{\cardA} y_{ime'} + \sum_{m\in\sM_j} \sum_{e'=0}^{e-1} x_{jme'} \leq 1 \quad (i,j)\in\sPrecCore,e\in\sSetminus{\sE}{0} \label{ct-see-precedence1}\\
    & r_{0k} = \sum_{i\in\sA}\sum_{m\in\sMi} b_{imk}x_{im0}  \qquad k\in\sR \label{ct-see-resource1}\\
    & r_{ek} = r_{e-1,k} + \sum_{i\in\sA}\sum_{m\in\sMi} b_{imk}(x_{ime}-y_{ime})  \;\; e\in\sE\setminus\{0,\cardA\}, k\in\sR \label{ct-see-resource2}\\
    & \sum_{i\in\sA}\sum_{m\in\sMi}\sum_{e\in\sSetminus{\sE}{\cardA}} w_{imk}x_{ime}\leq W_k  \quad k\in\sN \label{ct-see-nr-resources}\\
    & x_{ime}\in\{0,1\}  \qquad i\in\sA, m\in\sMi, e\in\sSetminus{\sE}{\cardA} \label{ct-see-dom-x}\\
    & y_{ime}\in\{0,1\}  \qquad i\in\sA, m\in\sMi, e\in\sSetminus{\sE}{0} \label{ct-see-dom-y}\\
    & s_e \geq 0  \qquad e\in\sE  \label{ct-see-dom-s}\\
    & 0 \leq  r_{ek} \leq B_k  \qquad e\in\sSetminus{\sE}{\cardA},k\in\sR. \label{ct-see-dom-r}
\end{align}
\end{subequations}
The objective~\eqref{ct-see-obj} minimizes the makespan. Constraints~\eqref{ct-ooe-first-event-start} and~\eqref{ct-ooe-event-order} impose that event numbers are nondecreasing in the time of occurrence starting with event zero at time zero. By constraints~\eqref{ct-see-event-time-def}, if an activity starts processing in mode $m$ at event $e$ and finishes processing in mode $m$ at event $f$, then event $f$ must be at least $p_{im}$ time units later than event~$e$. With~\eqref{ct-see-start-event} and~\eqref{ct-see-end-event}, it is ensured that each activity is assigned exactly one start and one end event. The set of constraints~\eqref{ct-see-precedence2} guarantees that an activity may not end at the same time or before it starts. The latter constraints can be equivalently formulated by a smaller set of constraints, however, computational experiences gained by \cite{Artigues2013} for the \ac{sm} case discourage from doing so. \eqref{ct-see-precedence1} enforce the given precedence relationships.  Constraints~\eqref{ct-see-resource1} and~\eqref{ct-see-resource2} implement the definition of the resource consumption variables~$r_{ek}$ in a recursive fashion. 
The non-renewable resource restrictions are implemented by~\eqref{ct-see-nr-resources}. Finally, \eqref{ct-see-dom-x}-\eqref{ct-see-dom-r} define the variable domains, where the upper bounds on $r_{ek}$ guarantee that renewable resources are not overloaded at any time.

For instances of \ac{mr} (i.e., with~$\sN=\emptyset$), the model \eqref{ct-see} above resembles the \ac{se}-based formulations of \cite{Chakrabortty2014} and \cite{Ozturk2020} in a slightly different but equivalent way. These minor differences are: (i) we additionally rule out a priori the possibility to start an activity at event~$\cardA$ or to finish an activity at event~$0$ by setting all~$x_{im\cardA}=y_{im0}:=0$, and (ii) \cite{Chakrabortty2014} track resource consumptions at the mode-level, which is not necessary. (In fact, an equivalent model without the use of auxiliary variables is readily available,
however, we keep these variables for a better readability and comparability.)

To show the incompleteness of the formulation defined in~\eqref{ct-see}, we next give a counterexample which leads to a schedule that is infeasible in \ac{mm}.

\begin{example}\label{ex:1}
Consider a project with two activities $i\in\sA:=\{1,2\}$ where activity 1 must precede activity 2, so $\sPrecCore:=\{(1,2)\}$. Suppose that two modes are given for each activity, e.g., $\sM_1=\sM_2:=\{1,2\}$ and the durations are $p_{11}=p_{21}:=1$ and $p_{12}=p_{22}:=2$. The activities consume only one renewable resource ($\sR:=\{1\}$) with $B_1:=1$ and all $b_{im1}:=1$.
\end{example}

Consider Solution A shown in Table~\ref{tab:ex-solns}, which is an optimal solution to model~\eqref{ct-see} in Example~\ref{ex:1}. Feasibility of the $x_{ime}$ and $y_{ime}$ values is easily checked by plugging them into constraints~\eqref{ct-see-start-event}-\eqref{ct-see-precedence1}. 

Moreover, it follows from~\eqref{ct-see-resource1} and~\eqref{ct-see-resource2} that
\begin{equation*}
r_{01} = 1  \leq B_1 \quad \text{ and } \quad  r_{11} = 1  \leq B_1 \quad \text{ and } \quad  r_{21} = 0  \leq B_1,
\end{equation*}
hence, the feasibility of the $r_{ek}$ values.
Regarding the $s_{e}$, constraint \eqref{ct-see-first-event-start} is trivial. Now, observe that in Solution A there only exist $i\in\sA,m\in\sMi$, and $e,f\in\sE$ with $e<f$ such that $x_{imk}=0$ or $y_{imk}=0$ (or both). This causes constraints~\eqref{ct-see-event-time-def} to collapse into the form
\[
s_f \geq s_e \quad \text{for all~$e,f\in\sEE$},
\]
which are already implied by~\eqref{ct-see-event-order}. As a result, all starting time variables $s_e$ can be set to zero and hence $z^\text{\ac{se}}=s_2=0$. However, Solution A is obviously an infeasible schedule for \ac{mm} because (i) the durations and capacity limits are ignored and (ii) the execution modes are chosen inconsistently.

\begin{table}[htb]
\caption{Two \ac{mm} solutions considered in Example~\ref{ex:1}}
\label{tab:ex-solns}
\bigskip
\renewcommand{\arraystretch}{0.5} 
\centering
\begin{tabular}{rrrr|rrr}
\toprule
& \multicolumn{3}{c}{Soln.\ A} & \multicolumn{3}{c}{Soln.\ B} \\
\cmidrule(lr){2-4}\cmidrule(lr){5-7}
 $e$        & 0  & 1  & 2                   & 0  & 1  & 2\\
\midrule
$x_{11e}$   & 0  & 0 & 0                    & \ccell{cellgray1}{1} & 0 & 0 \\
$y_{11e}$   & 0  & \ccell{cellgray1}{1} & 0  & 0  & \ccell{cellgray1}{1} & 0 \\[2ex]
$x_{12e}$   & \ccell{cellgray1}{1}  & 0 & 0  & 0  & 0 & 0 \\
$y_{12e}$   & 0  & 0 & 0                    & 0  & 0 & 0 \\[2ex]
$x_{21e}$   & 0  & 0 & 0  & 0  & \ccell{cellgray2}{1} & 0 \\
$y_{21e}$   & 0  & 0 & \ccell{cellgray2}{1} & 0  & 0 & \ccell{cellgray2}{1} \\[2ex]
$x_{22e}$   & 0  & \ccell{cellgray2}{1} & 0 & 0  & 0 & 0 \\
$y_{22e}$   & 0  & 0 & 0  & 0  & 0 & 0 \\
\midrule
$r_{e1}$    & 1 & 1 & 0 & 1 & 1 & 0 \\
$s_e$       & 0 & 0 & 0 & 0 & 1 & 2 \\
\bottomrule
\end{tabular}\\
\bigskip
\footnotesize \textit{Notes:} Solution A is an optimal solution in~\eqref{ct-see} but infeasible in \ac{mm}. Solution B is an optimal solution in both the model~\eqref{ct-see} including~\eqref{ct-see-vi1} and in~\ac{mm}.
\end{table}

Example~\ref{ex:1} identifies a serious flaw in the models presented by \cite{Chakrabortty2014} and \cite{Ozturk2020} as their models fail to ensure a consistent mode choice for each activity. However, we can correct this by adding to model~\eqref{ct-see}, e.g.,
\begin{equation}\label{ct-see-vi1}
1- x_{ime} \geq \sum_{m'\in\sSetminus{\sMi}{m}} \sum_{f\in\sSetminus{\sE}{0}} y_{im'f} \qquad i\in\sA,m\in\sMi, e\in\sSetminus{\sE}{\cardA}. \tag{\ref{ct-see}-MC}
\end{equation}
The constraints state that if activity $i$ starts operating in mode $m$ in event~$e$, than $i$ must not end in any mode other than $m$.%
\footnote{Notice that mode consistency could be equivalently established by interchanging the variables in~\eqref{ct-see-vi1} so that if $y_{imf}=1$, then $\sum_{m'\in\sSetminus{\sMi}{m}} \sum_{e\in\sSetminus{\sE}{\cardA}} x_{im'e}\leq 0$ for each $i\in\sA,m\in\sMi, f\in\sSetminus{\sE}{0}$.}

This establishes the missing link between the unique assignment of start and end events imposed by~\eqref{ct-see-start-event} and \eqref{ct-see-end-event}. Therefore, model~\eqref{ct-see} plus constraints~\eqref{ct-see-vi1} is a valid formulation of \ac{mm}.
Solution B in Table~\ref{tab:ex-solns} is a mode-consistent optimal schedule in Example~\ref{ex:1}, thus the minimum makespan increases to $z^\text{\ac{se}}=s_2=2$.

Alternatively, mode consistency can be established with the following smaller set of constraints:
\begin{equation}
\label{ct-see-mode-consistency-agg}
\sum_{e\in\sSetminus{\sE}{\cardA}} x_{ime} + \sum_{m'\in\sSetminus{\sMi}{m}}\sum_{f\in\sSetminus{\sE}{0}}  y_{im'f} \leq 1 \qquad i\in\sA, m\in\sMi. \tag{\ref{ct-see-vi1}-A}
\end{equation}

Validity of the latter inequalities is evident. Noting that, for any activity-mode combination, the left-hand side of~\eqref{ct-see-mode-consistency-agg} can be at most two because of~\eqref{ct-see-start-event} and \eqref{ct-see-end-event}, the inequality forbids solutions where activity $i$ ends operating in any mode other than the selected start mode. The model defined by~\eqref{ct-see} with~\eqref{ct-see-mode-consistency-agg} in place of \eqref{ct-see-vi1} is referred to as \textit{aggregate \ac{se} model}~(\acused{sea}\ac{sea}).

As a side note, the event time inequalities~\eqref{ct-see-event-time-def} can be stated in a stronger form; see \citep{Tesch2020}. In our preliminary computational tests, however, the \ac{se}-based models performed consistently worse with the stronger constraints. We therefore stick to the stated formulation.

\begin{remark}
Since Example~\ref{ex:1} has only two activities and one precedence relationship, this might appear as a simplistic way to show the incompleteness. However, there are examples involving more than two activities. For instance, consider the problem described in Example~\ref{ex:1} with two more double-mode activities $i=3$ and $i=4$ where activity~3 must precede activity~4. Letting $p_{31}=p_{41}=1$, $p_{32}=p_{42}=2$, and all $b_{3m1}=b_{4m1}=1$, it is easily checked that the minimum makespan produced by the \ac{se} model without and with mode consistency constraints is 0 and 4, respectively.
\end{remark}

\subsection{Revised start-end-event formulation}

Next, we present a new \ac{ct} that builds upon the ideas of \cite{Bianco2013,Bianco2017} and~\cite{Tesch2020} who studied improved \ac{se}-based models for the \ac{sm}. 

Define binary variables using the linear transformations:
\begin{align*}
    \varx_{ime} ={}& \sum_{\substack{e'\in\sSetminus{\sE}{\cardA}:\\e'\leq e}} x_{ime} \qquad i\in\sA,m\in\sMi,e\in\sSetminus{\sE}{\cardA}\\
    \vary_{ime} ={}& \sum_{\substack{e'\in\sSetminus{\sE}{0}:\\e'\leq e}} y_{ime} \qquad i\in\sA,m\in\sMi,e\in\sSetminus{\sE}{0}
\end{align*}
The interpretation of the new variables is slightly different because~$\varx_{ime}$ and~$\vary_{ime}$ state that activity~$i$ starts and, respectively, ends until event~$e$. Plugging the above transformations into~\eqref{ct-see} yields the so-called \acfi{rse} model for \ac{mm}:
\begin{subequations}
\label{ct-rse}
\begin{align}
        z^\text{RSEE} ={}& \min s_\cardA \label{ct-rsee-obj}\\
    \text{s.t. }&%
    s_0 = 0
    \label{ct-rsee-first-event-start}\\
    & s_e \leq s_{e+1} \quad e\in\sSetminus{\sE}{\cardA}
    \label{ct-rsee-event-order}\\
    & s_f \geq s_e + p_{im}(\vary_{imf} - \varx_{im,e-1}) \qquad i\in\sA,m\in\sMi,(e,f)\in\sEE 
    \label{ct-rsee-event-time-def}\\
    & \sum_{m\in\sMi} \varx_{im,\cardA-1} = 1 \qquad i\in\sA
    \label{ct-rsee-start-event}\\
    & \sum_{m\in\sMi} \vary_{im\cardA} = 1 \qquad i\in\sA
    \label{ct-rsee-end-event}\\
    & \varx_{ime}\leq\varx_{im,e+1} \qquad i\in\sA,m\in\sMi,e\in\sE: e<\cardA-1
    \label{ct-rsee-mono-x}\\
    & \vary_{ime}\leq\vary_{im,e+1} \qquad i\in\sA,m\in\sMi,e\in\sE:0<e<\cardA
    \label{ct-rsee-mono-y}\\
    & \vary_{ime} \leq \varx_{im,e-1} \quad i\in\sA,m\in\sMi,e\in\sSetminus{\sE}{0}
    \label{ct-rsee-start-before-end}\\
    & \sum_{m\in\sM_j} \varx_{jme} \leq \sum_{m\in\sMi} \vary_{ime} \quad (i,j)\in\sPrecCore,e\in\sSetminus{\sE}{\cardA} 
    \label{ct-rsee-precedences}\\
    &\sum_{i\in\sA}\sum_{m\in\sMi}b_{imk}(\varx_{ime} - \vary_{ime}) \leq B_k \qquad e\in\sSetminus{\sE}{\cardA},k\in\sR
    \label{ct-rsee-resources}\\
    & \sum_{i\in\sA}\sum_{m\in\sMi} w_{imk}\varx_{im,\cardA-1}\leq W_k  \qquad k\in\sN \label{ct-rsee-nr-resources}\\
    & \sum_{m'\in\sSetminus{\sMi}{m}} \vary_{im'\cardA} + \varx_{im,\cardA-1} \leq 1 \qquad i\in\sA, m\in\sMi
    \label{ct-rsee-mode-consistency}\\
    & \varx_{ime}\in\{0,1\}  \qquad i\in\sA, m\in\sMi, e\in\sSetminus{\sE}{\cardA} \label{ct-rsee-dom-x}\\
    & \vary_{ime}\in\{0,1\}  \qquad i\in\sA, m\in\sMi, e\in\sSetminus{\sE}{0} \label{ct-rsee-dom-y}\\
    & s_e \geq 0  \qquad e\in\sE, \label{ct-rsee-dom-s}
\end{align}
\end{subequations}
where we define~$\varx_{im,-1}=\vary_{im0}:=0$ for~$i\in\sA,m\in\sMi$.

\eqref{ct-rsee-obj}-\eqref{ct-rsee-event-order} are the same as in the original model.
The event time inequalities are given by~\eqref{ct-rsee-event-time-def}. They impose~$s_f\geq s_e + p_{im}$ if activity~$i$ ends in mode~$m$ until~$f$ but does not start in mode~$m$ until~$e-1$ with~$e<f$. Constraints~\eqref{ct-rsee-start-event} and~\eqref{ct-rsee-end-event} express that each activity must start in any mode until~$\cardA-1$ and, respectively, end in any mode until~$\cardA$. The new binary variables, by  definition, should form monotone sequences in every feasible solution, as written in~\eqref{ct-rsee-mono-x} and~\eqref{ct-rsee-mono-y}.
Constraints~\eqref{ct-rsee-start-before-end} state that activity~$i$ must start until~$e-1$ if it finishes at event~$e$. By~\eqref{ct-rsee-start-before-end}, each activity cannot end before it starts. The constraints~\eqref{ct-rsee-precedences} formulate the precedence requirements, i.e., any successor~$j$ of an activity~$i$ can only start until~$e$ if~$i$ finishes until~$e$. \eqref{ct-rsee-resources} and~\eqref{ct-rsee-nr-resources} define the capacity availability of the renewable and, respectively, non-renewable resources. The mode consistency constraints are given by~\eqref{ct-rsee-mode-consistency} and they stipulate that if an activity~$i$ starts in mode~$m$ until~$\cardA-1$, then it cannot end in any mode other than~$m$. 

Note that the two models~\eqref{ct-see} and~\eqref{ct-rse} are equivalent in terms of the LP polyhedron which follows, in general, from the non-singularity of the above transformations \citep{Artigues2017,Tesch2020}. Despite this fact, the \ac{rse} model provides a sparser constraint matrix than the original model (i.e., with less non-zero coefficients). Different authors reported independently that sparsity in \ac{sm} models is exploited successfully by current MIP solvers \citep{Bianco2013,Tesch2020}.

Finally, we notice that \cite{Tesch2020} introduced another event-based model for \acp{sm} that he calls the \emph{interval event-based formulation} (IEE). The author found out that the IEE model strictly dominates all other event-based \acp{ct} in terms of the LP relaxation. While this is a theoretically appealing feature of the model, the number of binary variables increases quadratically in the number of events. We therefore constrain our evaluation to the more compact event-based formulations wherein the number of binary variables increases only linearly in the events.


\subsection{On-off-event formulation}
\label{sec:ct-ooe}
In this section, we present a \ac{ct} for \ac{mm} that uses the notion of \acfp{oo}. An \ac{oo} \citep[see][]{Bowman1959,Kone2011} refers to an instant of time where an activity is either in process or is not being processed. Hence, there are as many events \acp{oo} as the number of activities. Let their index set be given by~$\sSetminus{\sE}{\cardA}$ and notice that~$\sEE=\{(e,f)\in\sSetminus{\sE}{\cardA}\times\sE:e<f\}$.

Define binary variables $z_{ime}$ that are equal to one if and only if activity~$i$ is being processed (or, \enquote{active}) in mode~$m$ at event $e$. Continuous variables $s_e, e\in\sSetminus{\sE}{\cardA},$ measure the point in time when event $e$ occurs, and one additional variable, $s_\cardA$, indicates the project makespan. With the help of auxiliary variables~$r_{ik}$ we keep track of the amount of capacity of a non-renewable resource~$k\in\sN$ that is consumed by an activity $i\in\sA$ during the project.
\ac{mm} can now be formulated as:
\begin{subequations}
\label{ct-ooe}
\begin{align}
    z^\text{OOE} ={}& \min s_\cardA \label{ct-ooe-obj}\\
    \text{s.t. }& s_0 = 0 \label{ct-ooe-first-event-start}\\
    & s_{e+1} \geq s_e \quad e\in\sSetminus{\sE}{\cardA} \label{ct-ooe-event-order}\\
    & s_f \geq s_e + p_{im}(z_{ime} - z_{im,e-1} + z_{im,f-1} - z_{imf} - 1) \nonumber\\
    & \hphantom{s_f \geq s_e + p_{im}(z_{ime}- z_{im,e-1} +} \qquad i\in\sA,m\in\sMi
    , (e,f)\in\sEE \label{ct-ooe-event-time-def}\\
    & \sum_{m\in\sMi} \sum_{e\in\sSetminus{\sE}{\cardA}} z_{ime} \geq 1 \qquad i\in\sA \label{ct-ooe-atleast-one-event}\\
    & \sum_{m'\in\sMi}\sum_{e'=0}^{e-1} z_{im'e'} \leq e\left(1 - (z_{ime}- z_{im,e-1})\right) \nonumber\\
    & \hspace{5cm} i\in\sA,m\in\sMi, e\in\sE\setminus\{0,\cardA\} \label{ct-ooe-contiguity-bw} \\
    & \sum_{m'\in\sMi}\sum_{e'=e}^{\cardA-1} z_{im'e'} \leq (\cardA - e)\left(1 - (z_{im,e-1} - z_{ime})\right)
    \nonumber\\
    & \hspace{5cm}  i\in\sA,m\in\sMi, e\in\sE\setminus\{0,\cardA\} \label{ct-ooe-contiguity-fw}\\
    & \sum_{m'\in\sM_j} \sum_{e'=0}^{e} z_{jm'e'} \leq (e + 1)(1 - z_{ime}) \nonumber\\ &\hspace{5cm}(i,j)\in\sPrecCore,m\in\sMi,e\in\sSetminus{\sE}{\cardA} \label{ct-ooe-precedence}\\
    & \sum_{i\in\sA}\sum_{m\in\sMi} b_{imk}z_{ime} \leq B_k \quad e\in\sSetminus{\sE}{\cardA}, k\in\sR \label{ct-ooe-resource}\\
    & \sum_{m\in\sMi}w_{imk}z_{ime} \leq r_{ik} \quad i\in\sA, e\in\sSetminus{\sE}{\cardA},k\in\sN \label{ct-ooe-nr-resource}\\
    & \sum_{i\in\sA} r_{ik} \leq W_k \quad k\in\sN \label{ct-ooe-nr-resource2}\\
    & z_{ime}\in\{0,1\}  \quad i\in\sA, m\in\sMi, e\in\sSetminus{\sE}{\cardA} \label{ct-ooe-dom-z}\\
    & r_{ik} \geq 0  \qquad i\in\sA,k\in\sN, \label{ct-ooe-dom-r} \\
    & s_e \geq 0  \qquad e\in\sE, \label{ct-ooe-dom-s}
\end{align}
\end{subequations}
where we use dummy variables~$z_{im,-1}=z_{im\cardA}:=0$ for each~$i\in\sA,m\in\sMi$.

\eqref{ct-ooe-obj}-\eqref{ct-ooe-event-order} are the same as in the \ac{se} model.  \eqref{ct-ooe-event-time-def} are the event time inequalities stating that if activity~$i$ starts in mode~$m$ at event~$e$ and ends in~$m$ at event~$f>e$, then~$s_f$ must be at least as large as~$s_e+p_{im}$.
Constraints~\eqref{ct-ooe-atleast-one-event} force each activity to be active at least once. Constraints~\eqref{ct-ooe-contiguity-bw} and~\eqref{ct-ooe-contiguity-fw} are referred to as \emph{contiguity constraints}. They implement the non-preemption requirement permitting only unimodal \emph{active sequences}~$z_{im0},z_{im1},\dots,z_{im\cardA}$ for~$i\in\sA,m\in\sMi$. It means that a particular activity-mode combination~$(i,m)$ may be active at one or more events but, in any case, processing starts at~$e$ iff $z_{ime} - z_{im,e-1}=1$ and ends at~$f$~iff $z_{im,f-1} - z_{imf}=1$ with~$(e,f)\in\sEE$. Constraints~\eqref{ct-ooe-precedence} enforce that the precedence relationships between activities are reflected in the event assignments. If activity $i$ starts at $e$ and must precede $j$, then $j$ must not be processed at or before~$e$.
Constraints~\eqref{ct-ooe-resource} implement the capacity limitations for renewable resources, while \eqref{ct-ooe-nr-resource} together with~\eqref{ct-ooe-nr-resource2} limit  the capacity of non-renewable resources. Finally, \eqref{ct-ooe-dom-z}-\eqref{ct-ooe-dom-s} define the variable domains.

Recent works on \ac{sm} models \citep[e.g.,][]{Nattaf2019, Tesch2020} give rise to the possibility to replace constraints~\eqref{ct-ooe-event-time-def}, \eqref{ct-ooe-contiguity-bw}, \eqref{ct-ooe-contiguity-fw}, and \eqref{ct-ooe-precedence} by stronger, i.e., \emph{dominating}, inequalities. For example, event time inequalities can be formulated as
\begin{equation}
    \label{ct-ooe-start-time-strong}
s_g \geq s_e + p_{im}(z_{imf} - z_{ime} - z_{img})
    \tag{\ref{ct-ooe-event-time-def}$^\prime$}
\end{equation}
for all~$i\in\sA,m\in\sMi$ and~$e,f,g\in\sE$ with~$e<f<g$, which means that if activity~$i$ is active in mode~$m$ at event~$f$ but inactive at~$e$ and~$g$, then~$s_g\geq s_e + p_{im}$ must hold. 

Dominating inequalities like~\eqref{ct-ooe-start-time-strong} bear the potential to produce tighter LP bounds. However, it has been shown for \acp{sm} \citep{Tesch2020} that all known stronger variants are still weak from a polyhedral point of view as they do not improve the optimal value of the LP relaxation. In an attempt to keep our model as small possible, we stick to the weaker inequalities in the stated model.

Notice the similarity of our stated \ac{oo} model to that of \cite{Chakrabortty2014} for the case~$\sN=\emptyset$.
Mode-consistency issues are not mentioned in their paper, hence, we question the merits of adding, e.g.,
\begin{equation}
\label{ct-ooe-mode-consistency}
\sum_{m'\in\sSetminus{\sMi}{m}}\sum_{f\in\sSetminus{\sE}{\cardA}} z_{im'f} \leq \cardA(1 - z_{ime}) \quad i\in\sA, m\in\sMi,e\in\sSetminus{\sE}{\cardA} \tag{\ref{ct-ooe}-MC}.
\end{equation}
One argument against employing~\eqref{ct-ooe-mode-consistency} is based on the contiguity constraints~\eqref{ct-ooe-contiguity-bw} and~\eqref{ct-ooe-contiguity-fw}. Consider a solution in which, say, two execution modes~$\mu_1,\mu_2\in\sMi$ are selected for some activity~$i\in\sA$. This solution implies two active sequences~$\sigma_1=(z_{i\mu_1 e_1},\dots,z_{i\mu_1 f_1})$ and~$\sigma_2=(z_{i\mu_2 e_2},\dots,z_{i\mu_2 f_2})$ with $(e_1,f_1)\in\sEE$ and $(e_2,f_2)\in\sEE$ marking the start/end of the activity-mode combination $(i,\mu_1)$ and $(i,\mu_2)$, respectively. The contiguity constraints forbid that~$e_1\neq e_2$ or~$f_1\neq f_2$ (or both). In the case~$e_1=e_2$ and~$f_1=f_2$, the solution implies tighter event time inequalities~\eqref{ct-ooe-event-time-def} leading, at most, to a larger objective function value. Therefore, either an optimal solution is mode consistent or mode consistency has no impact on the optimal makespan (in which case $\sigma_1$ or $\sigma_2$ can be chosen arbitrarily). 
This observation is in stark contrast to the \ac{se} model of Section~\ref{sec:ct-see} where choosing two or more modes can \enquote{undermine} the purpose of makespan-determining event time constraints.
Another argument against constraints~\eqref{ct-ooe-mode-consistency} suggests that they do \emph{not} help improving its LP bound. We show this fact below.

\begin{lem}\label{lem:zero-makespan}
The minimum makespan of the LP relaxation of the \ac{oo} model~\eqref{ct-ooe} with and without mode-consistency constraints~\eqref{ct-ooe-mode-consistency} is equal to zero.
\end{lem}
\begin{proof}
See~\ref{sec:appendix-proof}.
\end{proof}

However, it should be emphasized that mode-consistency constraints can be beneficial in other \ac{oo}-based formulations. As an example, let us consider a version of model~\eqref{ct-ooe} which consists of~\eqref{ct-ooe-obj}-\eqref{ct-ooe-atleast-one-event}, \eqref{ct-ooe-resource}-\eqref{ct-ooe-dom-s}, and

\begin{align}
\label{ct-ooe-contiguity-bw-aggr}
\sum_{m\in\sMi} \sum_{e'=0}^{e-1} z_{ime'} &\leq e\left(1 - \sum_{m\in\sMi}(z_{ime}- z_{im,e-1})\right) \;  i\in\sA, e\in\sE\setminus\{0,\cardA\} \tag{\ref{ct-ooe-contiguity-bw}$^\prime$} \\
\label{ct-ooe-contiguity-fw-aggr}
\sum_{m\in\sMi} \sum_{e'=e}^{\cardA-1} z_{ime'} &\leq (\cardA - e)\left(1 - \sum_{m\in\sMi}(z_{im,e-1} - z_{ime})\right) \; i\in\sA, e\in\sSetminus{\sE}{\cardA} \tag{\ref{ct-ooe-contiguity-fw}$^\prime$}\\
\sum_{m\in\sM_j} \sum_{e'=0}^{e} z_{jme'} & \leq (e + 1)\left(1 - \sum_{m\in\sMi} z_{ime}\right) \quad (i,j)\in\sPrecCore,e\in\sSetminus{\sE}{\cardA} \tag{\ref{ct-ooe-precedence}$^\prime$}.
\end{align}
We refer to this formulation as aggregate \ac{oo} model (\acused{ooa}\ac{ooa}).
Note that \ac{ooa} allows for multiple active sequences for an activity which may start/end at \emph{different} events (e.g., one could select $\sigma_1$ for~$(i,\mu_1)$ and~$\sigma_2$ for~$(i,\mu_2)$ with~$(e_1,f_1),(e_2,f_2)\in\sEE$ such that~$e_1\neq e_2$ and $f_1\neq f_2$). Solutions of this kind can be forbidden by means of constraints~\eqref{ct-ooe-mode-consistency}. We therefore always include them when evaluating the \ac{ooa} model.

\subsection{Flow-based formulation}
\label{sec:ct-fct}
\cite{Artigues2003} proposed to cast \ac{sm} as a network flow optimization problem defined on a simple directed graph with a node set~$\sV:=\sA\cup\{0,\terminalA\}$ that includes the set of activities~$\sA$,
a super-source ($i=0$), and a super-sink ($i=\cardA+1=:\terminalA$). These two nodes represent dummy activities with zero resource and time consumption standing for the start and end of the project.
The arcs of the graph are defined by a set~$\sPrec\subseteq\sV\times\sV$ such that there is one arc~$(i,j)\in\sPrec$ for each precedence relationship given in $\sPrecCore$ plus one arc $(0,i)$ for each non-dummy activity $i$ without predecessor and one arc $(i,\terminalA)$ for each non-dummy activity without successor.
Then, $G=(\sV,\sPrec)$ denotes a directed acyclic precedence graph, often referred to as \acfi{aon} graph.

The model in \citep{Artigues2003} uses two types of continuous variables to capture flows and starting times, and binary \emph{sequencing} variables. 
To extend this network flow model to \ac{mm}, we first reserve a dummy mode~$m=0$ for the dummy activities (i.e., $\sM_0=\sM_\terminalA:=\{0\}$) and define all dummy activity-mode combinations to consume zero time and zero capacity. Second, we introduce an additional set of binary mode-assignment variables, similar to \citep{Gnaegi2019,Ozturk2020}.

More precisely, we define a sequence variable $y_{ij}=1$ iff activity $i$ must complete before activity $j$ starts, a mode-assignment variable $x_{im}=1$ iff activity $i$ is assigned mode $m$, and an arc flow variable $f_{ijk} \geq 0$ to control the amount of capacity that is transferred from node $i$ (directly after $i$ ends) to node $j$ (directly before $j$ starts). Starting time variables are denoted by~$s_{i}\geq 0, i\in\sV$, where $s_\terminalA$ indicates the project makespan.

We use~$\sH$ to refer to the transitive hull of the \ac{aon} graph $G$. That means if a node $j\in\sV$ is reachable from another node $i\in\sV$ in $G$, then transitivity implies a precedence relationship between activity $i$ and $j$, and so $(i,j)\in\sH$. Clearly, only the pairs of activities $(i,j)$ that are \emph{not} contained in $\sH$ can potentially be executed concurrently.
Below, we state the given network flow problem as a nonlinear \ac{mip}:
\begin{subequations}
\label{fct}
\begin{align}
    z^\text{FCT} ={}& \min s_\terminalA \label{fct-obj}\\
    \text{s.t. }&%
    \sum_{m\in\sMi} x_{im} = 1 \qquad i\in\sV \label{fct-assign-one-mode}\\
    & s_i + y_{ij}\cdot\sum_{m\in\sMi} x_{im}p_{im} \leq s_j + \bigM(1-y_{ij})  \quad (i,j)\in\sV \times \sV \label{fct-starting-times-def-ozon}\\
    & y_{ij} + y_{ji} \leq 1 \qquad (i,j)\in\sV\times\sV:i<j \label{fct-seq-ordering}\\
    & y_{ij} + y_{jv} - y_{iv} \leq 1 \qquad i,j,v\in\sV: \text{\texttt{allDiff}}(i,j,v) \label{fct-seq-transitivity}\\
    & f_{ijk} \leq y_{ij}\cdot\sum_{m\in\sMi} \Tilde{b}_{imk}x_{im}  \quad i\in\sSetplus{\sA}{0},j\in\sSetplus{\sA}{\terminalA}: i\neq j; k\in\sR \label{fct-flow-upperbound-ozon}\\
    & \sum_{j\in\sV} f_{ijk} = \sum_{m\in\sM_i} \Tilde{b}_{imk}x_{im} \qquad i\in\sV, k\in\sR \label{fct-flow-balance-non-dummy-activities-1}\\
    & \sum_{i\in\sV} f_{ijk} = \sum_{m\in\sM_j} \Tilde{b}_{jmk}x_{jm}
    \qquad j\in\sV, k\in\sR \label{fct-flow-balance-non-dummy-activities-2}\\
    & f_{\terminalA jk} = \begin{cases}
    B_k& \text{if $j=0$}\\
    0 & \text{otherwise} \\
    \end{cases}\qquad j\in\sSetplus{\sA}{0}, k\in\sR \label{fct-flow-init}\\
    & f_{iik}=0 \qquad i\in\sV, k\in\sR \label{fct-no-loops}\\
    & y_{ij} = 1  \qquad (i,j)\in\sH  \label{fct-mandatory-precedences-1}\\
    & y_{ji} = 0  \qquad (i,j)\in\sH  \label{fct-mandatory-precedences-2}\\
    & y_{ii} = 0  \qquad i\in\sV  \label{fct-no-self-precedence}\\
    & \sum_{i\in\sA}\sum_{m\in\sMi} w_{imk}x_{im} \leq W_k \qquad k\in\sN \label{fct-nonrenewable-resource}\\
    & s_0 = 0  \label{fct-project-start}\\ 
    & s_i \geq 0 \qquad i\in\sV \label{fct-dom-s}\\
    & f_{ijk}\geq 0 \qquad (i,j)\in\sV\times\sV, k\in\sR\label{fct-dom-f}\\
    & x_{im}\in\{0,1\}  \qquad i\in\sV, m\in\sMi \label{fct-dom-x}\\
    & y_{ij}\in\{0,1\}  \qquad (i,j)\in\sV\times\sV, \label{fct-dom-y}
\end{align}
\end{subequations}
where~$\bigM$ is a sufficiently large number and
\[
\Tilde{b}_{imk}:= \begin{cases}
b_{imk}& \text{if~$i\in\sA$}\\
B_k& \text{if~$i\in\{0,\terminalA\}$}
\end{cases} \quad \text{ for~$i\in\sV,m\in\sMi$, and $k\in\sR$}.
\]

The objective function~\eqref{fct-obj} minimizes the makespan. Constraints~\eqref{fct-assign-one-mode} enforce that each activity is assigned exactly one mode, in particular, $x_{00}=x_{\terminalA 0}=1$ is implied. The nonlinear constraints~\eqref{fct-starting-times-def-ozon} couple starting times with sequence and mode-assignment decisions.
These constraints force any two activities $i$ and $j$ to be processed one after the other when $i$ passes a positive capacity flow to~$j$.
Constraints~\eqref{fct-seq-ordering}, \eqref{fct-seq-transitivity}, and \eqref{fct-no-self-precedence} assure a partial ordering of activities by forbidding $y_{ij}=y_{ji}=1$ and, respectively, by taking care of transitive precedence relationships.
Moreover, fixing the sequence variables according to~\eqref{fct-mandatory-precedences-1} and \eqref{fct-mandatory-precedences-2} enforces all pre-defined precedence relationships.
The nonlinear constraints~\eqref{fct-flow-upperbound-ozon} provide upper bounds on the resource flow along the $k$-th arc from node~$i$ to~$j$. \eqref{fct-flow-balance-non-dummy-activities-1} and \eqref{fct-flow-balance-non-dummy-activities-2} ensure flow conservation for all nodes.
To initialize the network, flow variables corresponding to arcs emanating from node $i=\terminalA$ are set as defined in~\eqref{fct-flow-init}. It means that the only positive flows leaving the sink go directly into the source node. \eqref{fct-no-loops} prohibit a positive flow along loops.
\eqref{fct-nonrenewable-resource} limit the usage of non-renewable resources. \eqref{fct-project-start} defines the project start at time zero, and \eqref{fct-dom-s}-\eqref{fct-dom-y} are the variable domains.

It should be mentioned that further mode consistency constraints are not needed here because, by \eqref{fct-assign-one-mode}, once an activity starts operating in an operational mode, it cannot change amid processing. The same holds for assignment-based models clearly.

The stated nonlinear model~\eqref{fct} has been proposed by \cite{Ozturk2020} for the case of \ac{mr}. The authors apply standard linearization techniques to eliminate nonlinear product terms. The resulting linear \ac{mip} has $\bigO(|\sV|^2 + |\sV|^2\dot|\sR|)$ additional continuous variables and defining inequalities.

In this paper, we consider two versions of the model~\eqref{fct}. For the first version, we replace~\eqref{fct-starting-times-def-ozon} and~\eqref{fct-flow-upperbound-ozon}, respectively, by
\begin{align}
s_i + \sum_{m\in\sMi} p_{im}x_{im} \leq{}& s_j + \bigMs(1-y_{ij})  \quad (i,j)\in\sV \times \sV : i \neq j \label{fct-starting-times-def-weak} \tag{\ref{fct-starting-times-def-ozon}$^{\prime}$}\\
\shortintertext{and}
f_{ijk} \leq{}& \bigMf y_{ij} \qquad i\in\sSetplus{\sA}{0},j\in\sSetplus{\sA}{\terminalA}:i\neq j; k\in\sR, \label{fct-flow-upperbound-weak} \tag{\ref{fct-flow-upperbound-ozon}$^{\prime}$}
\shortintertext{where we define}
\bigMs :={}& L_i + p^{\max}_i - E_j \qquad \text{for~$(i,j)\in\sV\times\sV$} \nonumber\\
\bigMf:={}& \min\{\Tilde{b}^{\max}_{ik},\Tilde{b}^{\max}_{jk}\} \qquad \text{for $i\in\sSetplus{\sA}{0},j\in\sSetplus{\sA}{\terminalA}:i\neq j;k\in\sR$} \nonumber
\end{align}
with  $p^{\max}_i:=\max\{p_{im} : m\in\sMi\}$ for~$i\in\sA$ and $\Tilde{b}^{\max}_{ik}:=\max\{\Tilde{b}_{imk}:m\in\sMi\}$ for~$i\in\sV,k\in\sR$.

Validity of inequalities \eqref{fct-starting-times-def-weak} and \eqref{fct-flow-upperbound-weak} is easy to verify, given the above definitions of~$\bigMs$ and~$\bigMf$. These two sets of constraints represent the original nonlinear inequalities in a more compact linear form than in the formulation of \cite{Ozturk2020} but they impose weaker upper bounds on the arc flows $f_{ijk}$. As an example, if $y_{ij}=1$, the right-hand side of~\eqref{fct-flow-upperbound-weak} evaluates to $\bigMf$. In this case, the right-hand side value of~\eqref{fct-flow-upperbound-ozon} equals $\sum_{m\in\sM_i} b_{imk}x_{im}$ which may be less than $\bigMf$, by definition. 
For this reason, we refer to this compact linear \ac{mip} as \acfi{fctw}. Note that \ac{fctw} is naturally linear and therefore more compact than the model obtained by linearizing~\eqref{fct}.

Our second model is an attempt to overcome the weak arc flow bounds present in both the model~\eqref{fct} and \ac{fctw}. It is based on a stronger version of the arc flow-bounding inequalities~\eqref{fct-flow-upperbound-ozon} given below.
\begin{lem}
The following arc flow-bounding inequalities
\begin{equation} 
\label{fct-flow-upperbound-strong}
f_{ijk} \leq y_{ij} \,  \ast \,  \min\left\{\sum_{m\in\sM_i} \Tilde{b}_{imk}x_{im},\sum_{m\in\sM_j} \Tilde{b}_{jmk}x_{jm}\right\}
\tag{\ref{fct-flow-upperbound-ozon}-S}
\end{equation}
for~$i\in\sSetplus{\sA}{0},j\in\sSetplus{\sA}{\terminalA},i\neq j,k\in\sR$ dominate inequalities~\eqref{fct-flow-upperbound-ozon}.
\end{lem}
\begin{proof}
The inequalities~\eqref{fct-flow-upperbound-strong} state that if activity~$i$ is executed in some mode~$\mu_i$ and activity~$j$ in~$\mu_j$, then the maximum flow permitted along the $k$-th arc from node~$i$ to~$j$ is determined by the activity-mode combination that consumes the lesser amount of resource~$k$, i.e., $\min\{\Tilde{b}_{i \mu_i k},\Tilde{b}_{j \mu_j k}\}$. Therefore, the inequality is valid. 
Since~\eqref{fct-flow-upperbound-strong} can be re-written as
\begin{multline*}
f_{ijk} \leq \min\left\{y_{ij}\cdot\sum_{m\in\sM_i} \Tilde{b}_{imk}x_{im},%
\; y_{ij}\cdot\sum_{m\in\sM_j} \Tilde{b}_{jmk}x_{jm}\right\} \\
\forall i\in\sSetplus{\sA}{0},j\in\sSetplus{\sA}{\terminalA}:i\neq j;k\in\sR,
\end{multline*}
dominance over~\eqref{fct-flow-upperbound-ozon} is evident, which completes the proof.
\end{proof}

Indeed, the inequalities \eqref{fct-flow-upperbound-strong} generalize the arc flow-bounding inequalities used in the network flow formulation of \cite{Kone2011} for \acp{sm}.
Standard linearization requires $\bigO(|\sV|^2\dot|\sR|)$ additional continuous variables and defining inequalities to eliminate product terms plus another $\bigO(|\sV|^2\dot|\sR|)$ inequalities to resolve the minimization operator. 
Consequently, if model~\eqref{fct} with~\eqref{fct-flow-upperbound-strong} instead of~\eqref{fct-flow-upperbound-ozon} is linearized, the resulting model is significantly larger than \ac{fctw} but also stronger. It is referred to as \acfi{fcts}.  Our computational experiments below examine the question whether or not it pays off in terms of solver performance to use the stronger but less compact model.

\section{Model enhancements}
\label{sec:enhancements}
\subsection{Time window restrictions}
\label{sec:tw}

It is straightforward to reduce the domain of starting time variables by imposing pre-computed time windows $[E_i,L_i]$ with $E_i$ and $L_i$ denoting, respectively, the earliest and latest possible start time of an activity~$i\in\sV$ in an optimal schedule. Time windows have been widely used for \acp{sm} e.g., to formulate additional constraints in \acp{ct} \citep{Kone2011,Tesch2020} or to lift various necessary constraints in \acp{dt}  \citep{Demassey2005}.
Time windows are easily computed via the well-known critical path method (CPM), where $[E_0,L_0]$ is typically set to~$[0,0]$.
\cite{Talbot1982} extends the CPM to cope with multiple modes by taking the minimum possible duration, $p_i^{\min}:=\min_{m\in\sMi}p_{im}$, for each activity~$i\in\sV$ in both forward and backward passes. 

Next, we apply the same idea to our \acp{ct} of Section~\ref{sec:models}. We add the subsequent constraints to the \ac{se} model~\eqref{ct-see} which basically impose~$s_e\in[E_i,L_i]$ if activity~$i$ starts operating in any mode at event~$e$, i.e.,
\begin{varsubequations}{\ref{ct-see}-TW}
\label{ct-see-tw}
\begin{align}
&E_i \sum_{m\in\sMi}\sum_{e'=0}^e x_{ime} \leq s_e\leq L_\terminalA \nonumber \\
& \qquad \qquad\qquad + (L_i  - L_\terminalA)\sum_{m\in\sMi} x_{ime} \qquad i\in\sA,e\in\sE\setminus\{0,\cardA\}.\\
\intertext{If activity $i$ completes mode~$m$ at $e$ the next inequalities impose that~$s_e\in[E_i+p_{im},L_i+p_{im}]$}
& \sum_{m\in\sMi}(E_i +p_{im})\sum_{e'=1}^e y_{ime} \leq s_e \leq L_\terminalA  \nonumber\\
& \quad \qquad\qquad + \sum_{m\in\sMi}(L_i+p_{im} - L_\terminalA) y_{ime} \qquad i\in\sA,e\in\sSetminus{\sE}{0}. \\
\intertext{Finally, we restrict the makespan to lie within the interval}
&E_\terminalA \leq s_\cardA \leq L_\terminalA.
\end{align}
\end{varsubequations}

In a similar fashion, we add to the \ac{oo}-based model~\eqref{ct-ooe}
\begin{varsubequations}{\ref{ct-ooe}-TW}
\label{ct-ooe-tw}
\begin{align}
& E_i \sum_{m\in\sMi} z_{ime} \leq s_e   \nonumber\\
& \hphantom{E_i \sum_{m\in\sMi} z_{ime}} \leq L_\terminalA + (L_i - L_\terminalA) \sum_{m\in\sMi} z_{ime} \quad i\in\sA,e\in\sE\setminus\{0,\cardA\},\\
& \sum_{m\in\sMi} (E_i +p_{im}) (z_{im,e-1}-z_{ime}) \leq s_e \leq  L_\terminalA  \nonumber \\
& \; + \sum_{m\in\sMi}(L_i + p_{im}-L_\terminalA)(z_{im,e-1}-z_{ime}) \quad i\in\sA,e\in\sSetminus{\sE}{\cardA}:e>1, \\
&E_\terminalA \leq s_\cardA \leq L_\terminalA.
\end{align}
\end{varsubequations}


In the network flow models \ac{fctw} and \ac{fcts}, we replace constraints~\eqref{fct-project-start} and \eqref{fct-dom-s} by
\begin{equation}
E_i \leq s_i \leq L_i \quad i\in\sV. \label{fct-tw} \tag{\ref{fct}-TW}
\end{equation}

\subsection{Enforcing sequential processing for incompatible activities}

Incompatible activities are a well-known concept used in \ac{sm} to compute lower bounds on the makespan \citep{Stinson1978,Olaguibel1993,Klein1999}.
Loosely speaking, two activities are said to be incompatible if they cannot be processed in parallel due to scarce capacity of a common resource.
\cite{Gnaegi2019} suggest a set of redundant constraints for their assignment-based models that exploits (without naming it) incompatible activities, more precisely, activity-mode combinations. However, the applicability of these redundant constraints is not restricted to assignment-based models as we show next.

We denote an activity-mode combination by~$u=(i_u,m_{i_u})$ with $i_u\in\sA,m_{i_u}\in\sM_{i_u}$.
An activity-mode combination $(u,v)$ is referred to as an \emph{incompatible pair} if they share a common bottleneck resource, i.e., if $b_{i_u m_{i_u}k} + b_{i_v m_{i_v} k} > B_k$ for some~$k\in\sR$. Let~$\sS$ denote the set of all distinct incompatible pairs.
\cite{Gnaegi2019} identify a sufficient condition for enforcing sequential processing of an incompatible pair, namely if $i_u$ is executed in $m_{i_u}$ and $i_v$ in $m_{i_v}$. Now, this information can be exploited in the network flow model because only assignment and sequence variables are involved.
Therefore, the constraints
\begin{equation}
\label{VI-enforce-seq}
y_{i_u i_v} + y_{i_v i_u} \geq x_{i_u m_{i_u}} + x_{i_v m_{i_v}} - 1 \qquad (u,v)\in \sS \tag{\ref{fct}-RC}
\end{equation}
 can be added to model~\eqref{fct} without loss of optimality. Note that this idea easily extends to the more general notion of incompatible \emph{sets} \citep[see also][]{Olaguibel1993}. However, \cite{Gnaegi2019} reported decreasing computational performance for such cases, which they trace back to increasing numbers of constraints.

\subsection{Variable fixing}
\label{sec:oo-e-vf}

Deducing eliminable activity-end assignments is a popular reduction technique for event-based models~\citep{Kone2011,Tesch2020}. Let~$\sAncestors{i}$ and~$\sDescendants{i}$ denote, respectively, the number of ancestors and descendants of activity $i\in\sA$ in the \ac{aon} graph~$\Gbar=(\sA,\sPrecCore)$. Since the number of events considered in each event-based model of the previous section is at least as large as the number of activities and by noting that every activity must be processed in exactly one mode, it is without loss of optimality to eliminate all assignments of an activity~$i$ to any start event prior to~$\sAncestors{i}$ and not at or later than~$\cardA -\sDescendants{i}$. Likewise, activity~$i$ can neither finish at or earlier than~$\sAncestors{i}$ nor later than~$\cardA -\sDescendants{i}$. Formally, we can fix the variables $x_{ime}, y_{ime}$, and $z_{ime}$ as follows.
For the \ac{se} model~\eqref{ct-see}, we get
\begin{varsubequations}{\ref{ct-see}-VF}
\label{ct-see-vf}
\begin{align}
&x_{ime} = 0 \quad i\in\sA, m\in\sMi, e\in\sSetminus{\sE}{\cardA}:  e < \sAncestors{i} \text{ or } e\geq\cardA -\sDescendants{i} \\
&y_{ime} = 0 \quad i\in\sA, m\in\sMi, e\in\sSetminus{\sE}{0}: e\leq \sAncestors{i} \text{ or } e > \cardA-\sDescendants{i}.
\end{align}
\end{varsubequations}
Similarly, we get for the \ac{oo} model~\eqref{ct-ooe}:
\begin{align}
\label{ct-ooe-vf}
& z_{ime} = 0 \quad i\in\sA, m\in\sMi, e\in\sSetminus{\sE}{\cardA} : e < \sAncestors{i} \text{ or } e \geq \cardA -\sDescendants{i}. \tag{\ref{ct-ooe}-VF}
\end{align}


\section{Computational experiments}
\label{sec:experiments}
The main objective of this section is to provide a comparative study of the \acp{ct} discussed in this paper using an off-the-shelf \ac{mip} solver.%
Our experimental design is divided into two parts. The first part (Sections~\ref{sec:mm-experiments} and~\ref{sec:mm-results}) refers to experiments carried out in order to evaluate the general performance of the various models. In the second part (Sections~\ref{sec:mn-experiments} and~\ref{sec:mn-results}), we address the impact of an increasing number of modes on the computational complexity of the various models.
We do not intend to repeat the comparison between \acp{ct} and \acp{dt}, but we refer to existing studies \citep{Kone2011,Gnaegi2019,Ozturk2020} and to our introductory discussion in Section~\ref{sec:intro} instead.

All models used in our tests are implemented in Java\footnote{Find our implementation at \href{https://github.com/ddotsdot/project-scheduling-code}{https://github.com/ddotsdot/project-scheduling-code}} and we use SCIP\footnote{\href{https://www.scipopt.org}{https://www.scipopt.org}} (v8.0.0) as general-purpose \ac{mip} solver via a non-commercial declarative modelling API \citep{jdecor} which utilizes Google's OR Tools\footnote{\href{https://developers.google.com/optimization}{https://developers.google.com/optimization}} (v9.3.0). Our code employs the open-source programming library JGraphT\footnote{\href{https://jgrapht.org/}{https://jgrapht.org/}} \citep{jgrapht}, which provides useful graph-theoretical data structures and algorithms.  With a time limit of 300 seconds, all tests ran on a virtual machine with an Intel Xeon E5-2650 v2 at 2.60 GHz and 119 GB of RAM.

\subsection{Experiments with \ac{mm} instances}
\label{sec:mm-experiments}
We compare the following models:
\begin{description}
 \newcommand\mywidth{OOE-TW-VF-A}
\item[\myitem{\ac{se}-TW-VF}{\mywidth}] model~\eqref{ct-see} + \eqref{ct-see-vi1} + \eqref{ct-see-tw} + \eqref{ct-see-vf}
\item[\myitem{\ac{sea}-TW-VF}{\mywidth}] model~\ac{sea} + \eqref{ct-see-tw} + \eqref{ct-see-vf}
\item[\myitem{\ac{rse}}{\mywidth}] model~\eqref{ct-rse}
\item[\myitem{\ac{oo}-TW-VF}{\mywidth}] model~\eqref{ct-ooe} + \eqref{ct-ooe-tw} + \eqref{ct-ooe-vf}
\item[\myitem{\ac{ooa}-TW-VF}{\mywidth}] model~\ac{ooa} + \eqref{ct-ooe-tw} + \eqref{ct-ooe-vf} + \eqref{ct-ooe-mode-consistency}
\item[\myitem{\ac{fctw}-TW}{\mywidth}] model~\eqref{fct} + \eqref{fct-tw}
\item[\myitem{\ac{fcts}-TW}{\mywidth}] model~\eqref{fct} + \eqref{fct-tw}
\item[\myitem{\ac{fctw}-TW-RC}{\mywidth}] model~\eqref{fct} + \eqref{fct-tw} + \eqref{VI-enforce-seq}
\end{description}
The models listed above are benchmarked against two models from the literature: (i) the assignment-based formulation as defined in \cite{Gnaegi2019}, referred to as \acfi{ctab}, and (ii) the linear network flow formulation as defined in \cite{Ozturk2020}. The latter model, which we refer to as \acused{fcto}\ac{fcto}, is extended by additional constraints to incorporate non-renewable resources.
\cite{Gnaegi2019} report that \ac{ctab}, which is enhanced by additional constraints of the form~\eqref{VI-enforce-seq} and by fixing some of the resource-assignment variables, performs best among all model variants evaluated in their computational study. It therefore qualifies as a relatively strong competitor.

Two sets of \ac{mm} benchmark problems from the PSPLIB \citep{Kolisch1997} and one set from the MMLIB \citep{Peteghem2014} with 1545 instances in total build the basis for our tests. The data sets are called, respectively, \texttt{c15}, \texttt{j20}, and \texttt{J50}.\footnote{The first two data sets can be downloaded at \href{http://www.om-db.wi.tum.de/psplib/getdata_mm.html}{http://www.om-db.wi.tum.de/psplib/getdata\textunderscore mm.html} and the third set at  \href{https://www.projectmanagement.ugent.be/research/data}{https://www.projectmanagement.ugent.be/research/data}.}
Table~\ref{tab:instchar} shows problem size-related characteristics and the values of overall minimum/maximum resource demand ($b_{\min}$/$b_{\max}$) and resource capacity ($B_{\min}$/$B_{\max}$). 
We observe that these instances are characterized by rather low-valued demands and availabilities related to the renewable resources. This fact could be a competitive advantage for the \ac{ctab} model. For  a fairer and more meaningful comparison, we have derived three new data sets, denoted \texttt{c15\textunderscore d}, \texttt{j20\textunderscore d}, and \texttt{J50\textunderscore d}, by scaling up all renewable resource demands and availabilities with a constant factor $\delta:=10$.  

\begin{table}
    \caption{Characteristics of the benchmark instance sets \texttt{c15}, \texttt{j20}, and \texttt{J50}}
    \label{tab:instchar}
    \bigskip
    \centering
    \small\setlength{\tabcolsep}{8pt}
    \begin{tabular}{c|cccccccc}
    \toprule
    set & $\cardA$ & $|\sMi|$ & $|\sR|$ & $|\sN|$ & $b_{\min}$ & $b_{\max}$ & $B_{\min}$ & $B_{\max}$ \\
    \midrule
    \texttt{c15} & 15 & 3 & 2 & 2 & 1 & 10 & 5 & 45 \\
    \texttt{j20} & 20 & 3 & 2 & 2 & 1 & 10 & 5 & 50 \\
    \texttt{J50} & 50 & 3 & 2 & 2 & 1 & 10 & 15 & 108 \\
    \bottomrule
    \end{tabular}
\end{table}

We are aware of the fact that changing the relation between resource availability and demand in a particular instance can have a non-negligible increasing or even decreasing effect on the time an exact solution method would need to find an optimal schedule.
Consider, for example, the quantity referred to as \emph{resource strength} (RS) \cite{Kolisch1995} that relates the demand for a particular resource to its available capacity.
RS was invented in order to indicate the computational tractability of \ac{sm} instances measured in terms of solution times \citep[other measures are suggested in][]{Eynde2022}. It has been shown experimentally \citep{Reyck1996} that computational tractability of \ac{sm} instances is a bell-shaped function of RS; please consult \cite{Elmaghraby1980,Kolisch1995,Reyck1996} and to \cite{Peteghem2014, Eynde2022} about more details.
However, it is easy to show that if, for all~$k\in\sR$, we multiply all input parameters related to resource~$k$ (i.e., all $b_{imk}$ and $B_k$) with the same positive factor~$\delta$, then the $k$-th RS, as defined in \citep{Kolisch1995}, remains unchanged.
As a result, scaling instances in this way should not affect unintentionally the tractability of the original instances.

\subsection{Experiments with \ac{mn} instances}
\label{sec:mn-experiments}

Due to the inherent \ac{ma}, one would expect that a higher number of modes will ceteris paribus decrease the computational tractability of the overall scheduling problem. Recall that the \ac{ma} is a feasibility problem inherent in \acp{mm}. It can be stated as the problem to decide whether there exists a vector $(m(i))_{i\in\sA}$ of feasible mode assignments~$m(i)$ for each activity~$i\in\sA$ such that
\begin{itemize}
    \item each activity is assigned exactly one mode
    \item total non-renewable capacity consumption is within the given limits.
\end{itemize}
In order to examine the impact of an increasing number of modes in isolation, this section focuses on \ac{mn} benchmark problems that we generated by modifying three sets of \ac{mm} instances called  \texttt{m2}, \texttt{m4} and \texttt{m5} from the PSPLIB (1594 instances in total). We removed all renewable resources from each instance, that is, we set $\sR:=\emptyset$. The modified data sets are referred to as \texttt{m2N}, \texttt{m4N}, and \texttt{m5N}. Table~\ref{tab:instchar2} summarizes their characteristics. We compare the event-based models SEE-TW-VF, RSEE, and OOE-TW-VF with the strong flow-based model (\ac{fcts}-TW) and the assignment-based model (\ac{ctab}).

\begin{table}
    \caption{Characteristics of \ac{mn} test instances \texttt{m2N}, \texttt{m4N}, and \texttt{m5N}}
    \label{tab:instchar2}
    \bigskip
    \centering
    \small\setlength{\tabcolsep}{8pt}
    \begin{tabular}{c|cccccccc}
    \toprule
    set & $\cardA$ & $|\sMi|$ & $|\sR|$ & $|\sN|$ & $w_{\min}$ & $w_{\max}$ & $W_{\min}$ & $W_{\max}$ \\
    \midrule
    \texttt{m2N} & 16 & 2 & 0 & 2 & 1 & 10 & 19 & 127 \\
    \texttt{m4N} & 16 & 4 & 0 & 2 & 1 & 10 & 16 & 128 \\
    \texttt{m5N} & 16 & 5 & 0 & 2 & 1 & 10 & 20 & 139 \\
    \bottomrule
    \end{tabular}\\
\bigskip
\footnotesize \textit{Note:} The benchmark sets are derived from the PSPLIB data sets \texttt{m2}, \texttt{m4}, and \texttt{m5}, respectively.
\end{table}

\subsection{Results of the \ac{mm} experiments}
\label{sec:mm-results}

Our experiments concentrate on the relative performance in terms of integer solving. Regarding LP relaxations, we refer to our discussion in Section~\ref{sec:intro} and point the reader to similar investigations performed by~\citep{Artigues2003,Demassey2005,Kone2011,Artigues2013} for \ac{sm}.

\csvnames{my names}{benchmark=\benchmark,model=\model,feas=\feas,opt=\opt,best=\best,dev=\dev,cpu=\cpu, vars=\vars,cons=\cons}
\begin{table}[b!]
\caption{Numerical results for the benchmark data sets \texttt{c15} and \texttt{j20}}
\label{tab:results-PSPLIB}
\csvreader[
    my names,
    centered tabular=llrrrrrrr,
    respect underscore=true,
    before reading=\footnotesize,
    table head=%
    \toprule
    \thead{Data \\ Set} & \thead{Model \\ \\} & \thead{Feas \\ (\#)} & \thead{Opt \\ (\#)} & \thead{Best \\ (\#)} & \thead{$\Delta z$ \\ (\%)} & \thead{CPU \\ (s)} & \thead{Vars \\ (\#)} & \thead{Cons \\ (\#)}\\
    \midrule,
    table foot=\bottomrule
    ]%
    {data.csv}{}{%
    \benchmark  & \model & \feas & \opt & \best & \dev & \cpu & \vars & \cons
}
\vskip-1ex\centering\footnotesize \textit{Notes:} \texttt{c15} and \texttt{j20} are comprised of 551 and 554 instances; time limit = 300 sec.
\end{table}

\begin{table}[htb]
\caption{Numerical results for the benchmark data sets \texttt{J50}}
\label{tab:results-MMLIB}
\csvreader[
    my names,
    centered tabular=llrrrrrrr,
    respect underscore=true,
    before reading=\footnotesize,
    table head=%
    \toprule
    \thead{Data \\ Set} & \thead{Model \\ \\} & \thead{Feas \\ (\#)} & \thead{Opt \\ (\#)} & \thead{Best \\ (\#)} & \thead{$\Delta z$ \\ (\%)} & \thead{CPU \\ (s)} & \thead{Vars \\ (\#)} & \thead{Cons \\ (\#)}\\
    \midrule,
    table foot=\bottomrule
    ]%
    {dataJ50.csv}{}{%
    \benchmark  & \model & \feas & \opt & \best & \dev & \cpu & \vars & \cons
}
\vskip-1ex\centering\footnotesize \textit{Notes:} \texttt{J50} is comprised of 540 instances; time limit = 300 sec.
\end{table}

Tables~\ref{tab:results-PSPLIB}-\ref{tab:results-MMLIB-d} should be read as follows:  columns \enquote{Feas}, \enquote{Opt}, and \enquote{Best}  keep count of the number of times the solver was successful in finding a feasible solution, a proven optimal integer solution, and a solution with least makespan among those found. The last four columns show the percentage deviation from the least makespan found which is averaged over all available (but not provably optimal) solutions, the mean computation time in seconds, the mean number of variables, and the mean number of constraints. The models are ranked by Feas, Opt, and Best (in this order).

We start with a comparative analysis of the results for the data sets \texttt{c15}, \texttt{j20}, and \texttt{J50} depicted in Tables~\ref{tab:results-PSPLIB} and~\ref{tab:results-MMLIB}. 
It is obvious that the solver was able to find at least one feasible solution within the given time limit for all models in all instances.
Regarding exact solving, the weak network flow model (\acused{fct}\ac{fct}-TW-W) leads the field with about 88\% (80\%, 31\%) of the \texttt{c15} (\texttt{j20}, \texttt{J50}) instances solved to proven optimality. Also, note that the strong network flow model (\ac{fct}-S-TW) is inferior to the more compact flow models presented in this paper (\ac{fct}-W-x) and \ac{fcto}. It is therefore evident that the cost of linearizing the nonlinear constraints~\eqref{fct-flow-upperbound-strong} by means of standard techniques outweighs the benefits of linking flow and sequencing variables through stronger inequalities.

The results depicted for the event-based formulations are surprising for two reasons. First, the solver fails to prove optimality in all instances with the exception of \ac{rse} in case of the smallest instance set (\texttt{c15}). As to the two~\ac{oo}-based models, this result is in stark contrast to the insights gained by \cite{Kone2011} in the single-mode setting (therein, it is observed that the preprocessed \ac{oo}-based model can be solved to optimality in nearly as many \ac{sm} instances of the data set \enquote{KSD15\textunderscore d} as the flow-based model; it even outperforms all other competitors for the data set \enquote{PACK\textunderscore d}). Second, with increasing problem size the \ac{rse} model performs worse than most of the considered event-based alternatives. This contrasts the computational experiences of \cite{Tesch2020}, where \ac{rse} using CPLEX as reference MIP solver outperforms all other event-based models in his test bed.
\ac{ctab} ranks in the midfield with respect to all criteria.

\begin{table}[bt!]
\caption{Numerical results for the derived benchmark data sets \texttt{c15\textunderscore d} and \texttt{j20\textunderscore d}}
\label{tab:results-PSPLIB-d}
\csvreader[
    my names,
    centered tabular=llrrrrrrr,
    respect underscore=true,
    before reading=\footnotesize,
    table head=%
    \toprule
    \thead{Data \\ Set} & \thead{Model \\ \\} & \thead{Feas \\ (\#)} & \thead{Opt \\ (\#)} & \thead{Best \\ (\#)} & \thead{$\Delta z$ \\ (\%)} & \thead{Time \\ (s)} & \thead{Vars \\ (\#)} & \thead{Cons \\ (\#)}\\
    \midrule,
    table foot=\bottomrule
    ]%
    {data_d.csv}{}{%
    \benchmark  & \model & \feas & \opt & \best & \dev & \cpu & \vars & \cons
}
\vskip-1ex\centering\footnotesize \textit{Notes:} \texttt{c15\textunderscore d} and \texttt{j20\textunderscore d} are comprised of 551 and 554 instances; time limit = 300 sec.
\end{table}

\begin{table}[htb]
\caption{Numerical results for the benchmark data sets \texttt{J50\textunderscore d}}
\label{tab:results-MMLIB-d}
\csvreader[
    my names,
    centered tabular=llrrrrrrr,
    respect underscore=true,
    before reading=\footnotesize,
    table head=%
    \toprule
    \thead{Data \\ Set} & \thead{Model \\ \\} & \thead{Feas \\ (\#)} & \thead{Opt \\ (\#)} & \thead{Best \\ (\#)} & \thead{$\Delta z$ \\ (\%)} & \thead{CPU \\ (s)} & \thead{Vars \\ (\#)} & \thead{Cons \\ (\#)}\\
    \midrule,
    table foot=\bottomrule
    ]%
    {dataJ50_d.csv}{}{%
    \benchmark  & \model & \feas & \opt & \best & \dev & \cpu & \vars & \cons
}
\vskip-1ex\centering\footnotesize \textit{Notes:} \texttt{J50\textunderscore d} is comprised of 540 instances; time limit = 300 sec.
\end{table}

Considering Table~\ref{tab:results-PSPLIB-d} and~\ref{tab:results-MMLIB-d}, the above statements with respect to the relative performance between the flow-based models and the event-based models tend to hold  in the same way for the derived data sets \texttt{c15\textunderscore d} \texttt{j20\textunderscore d}, and \texttt{J50\textunderscore d}. However, as expected, the performance of \ac{ctab} drops drastically. While the solver was able to find at least one feasible solution in almost 100\% (61\%) of the small- and medium-sized (large-sized) instances that are characetrized by a lower range of capacity demands and availabilities, this fraction falls to 72\% (\texttt{c15\textunderscore d}), 54\% (\texttt{j20\textunderscore d}), and less than 1\% (\texttt{J50\textunderscore d}), which is far below the respective fractions achieved by the lowest-performing event-based model. Clearly, this result is due to the fact mentioned in Section~\ref{sec:intro} that the size of \ac{ctab} quickly increases by several orders of magnitude as the range of capacity-related parameter values increases (according to Table~\ref{tab:results-MMLIB-d} there are tens of thousands of variables and constraints on average). Nevertheless, it should be mentioned that in those instances where the solver is able to find feasible solutions for \ac{ctab}, the solver can often prove optimality within the time limit, e.g., in 203 (121) out of 394 (298) cases for \texttt{c15\textunderscore d} (\texttt{j20\textunderscore d}). But keep in mind that we used a rather moderate factor ($\delta=10$) to scale up the resource requirements and capacities.

\begin{table}[t!]
\caption{Numerical results for the benchmark data sets \texttt{m2N}, \texttt{m4N}, and \texttt{m5N}}
\label{tab:results-mxN}
\csvreader[
    my names,
    centered tabular=llrrrrrrr,
    respect underscore=true,
    before reading=\footnotesize,
    table head=%
    \toprule
    \thead{Data \\ Set} & \thead{Model \\ \\} & \thead{Feas \\ (\#)} & \thead{Opt \\ (\#)} & \thead{Best \\ (\#)} & \thead{$\Delta z$ \\ (\%)} & \thead{CPU \\ (s)} & \thead{Vars \\ (\#)} & \thead{Cons \\ (\#)}\\
    \midrule,
    table foot=\bottomrule
    ]%
    {data_mxN.csv}{}{%
    \benchmark  & \model & \feas & \opt & \best & \dev & \cpu & \vars & \cons
}
\vskip-1ex\centering\footnotesize \textit{Notes:} \texttt{m2N}, \texttt{m4N}, and \texttt{m5N} are comprised of 481, 555, and 558 instances, respectively; time limit = 100 sec.
\end{table}

\subsection{Results of the \ac{mn} experiments}
\label{sec:mn-results}

Table~\ref{tab:results-mxN} depicts the numerical results for the benchmark instances \texttt{m2N}, \texttt{m4N}, and \texttt{m5N}. The time limit was set to 100 seconds. 
The numbers reveal a significant gap in computational performance between the flow- and assignment-based models on the one hand and the three event-based models on the other hand. While the solver was able to find an optimal solution in less than a second for all data sets when solving \ac{ctab} or FCT-S-TW,  optimality could never be proven within the time limit in case of the event-based models (apart from one single instance). 

It can further be seen from Table~\ref{tab:results-mxN} that, for all event-based models, solver performance decreases (e.g., in terms of the number of feasible and best solutions found) when the number of modes increases. In contrast, solver performance is virtually insensitive in case of \ac{ctab} and FCT-S-TW for the tested instances.

Note that it is not too surprising that \ac{ctab} and FCT-S-TW perform equally well overall because, for \ac{mn} instances, both represent the inherent \ac{ma} in \emph{essentially} the same way, namely
\begin{align*}
     \sum_{m\in\sMi} x_{im} ={} &  1 \qquad i\in\sA \\
     \sum_{i\in\sA}\sum_{m\in\sMi} w_{imk}x_{im} \leq {} &  W_k \quad k\in\sN\\
     x_{im}\in \{0,1\} &   \qquad i\in\sA, m\in\sMi.
\end{align*}

This observation suggests the conjecture that overall competitiveness of the \ac{mm} formulations discussed in this paper is closely connected to the way how mode choices are modelled.

\subsection{Summary}
Based on the results of the previous subsections, we summarize our main insights as follows:
\begin{enumerate}
    \item \textit{Flow-based models}
    \begin{itemize}
    \item All network flow models (FCT-x) outperform the event-based models on almost all tested instances
    \item Among flow-based models, however, the weaker ones (FCT-W-x) consistently outperform the stronger ones because tightening the bounds on flow variables entails an expensive linearization
    \end{itemize}
\item \textit{Event-based models}
    \begin{itemize}
    \item All event-based models can compete with other model types only in terms of the ability to find any feasible solution but not in terms of the quality the integer solutions that were found
    \item Among event-based models, none consistently outperforms all other ones
    \item For \ac{se}-based models, \ac{se}-A consistently outperforms \ac{se}
    \item \ac{rse} rarely attains optimality, overall performance is rather below-average 
    \item For \ac{oo}-based models, neither \ac{oo} nor  \ac{oo}-A consistently outperfroms the other one
    \end{itemize}
\item \textit{Assignment-based model}
    \begin{itemize}
    \item Has decent overall performance on small and medium-sized instances
    \item Shows poor performance when renewable capacities are large
\end{itemize}
\end{enumerate}


\section{Concluding remarks}
\label{sec:conclusion}

In this paper, we reviewed compact continuous-time formulations for \acf{mm}.
We corrected a serious flaw in an existing model that uses \acfp{se} by formulating a set of so-called mode-consistency constraints, and we discussed merits of such constraints in a model that uses \acfp{oo}. 
In addition, we formulated a revised \ac{se} model previously proposed for the single-mode setting with promising results.
Then, we reconsidered a network flow model and provided two new variants that differ in the manner of bounding flow variables.
Along with the corrected and new models, we presented a couple of simple but usually effective model enhancements that are commonly used in the project scheduling literature.
We conducted extensive and fair computational experiments using established benchmark problems from the literature and newly created instances. The latter are characterized by a higher range of requirements and capacities with respect to the renewable resources. 
Finally, we examined the impact of an increasing number of modes on solution times.

In conclusion, our numerical results show a numerical dominance of the network flow formulations over the event-based models. While previous works addressing compact \ac{mip} formulations for \acp{mm} consider only instances up to 30 jobs, our network flow models make it possible to find optimal schedules in 31\% of the instances with 50 jobs, which on its own advances this field. Our experiments also give rise to claim that event-based formulations appear less competitive in multi-mode settings because of the way the inherent mode assignment problem is formulated.

However, compact \ac{mip} models have their limitations, e.g., when it comes to efficiently tackle problems with much more than 50 jobs. This creates room for future work. We see one opportunity in developing tailored (heuristic or exact) solution approaches exploiting the fact uncovered in this paper that mode assignments represent a computationally easier part of the overall scheduling problem.


\appendix
\section{Proof of Lemma~\ref{lem:zero-makespan}}
\label{sec:appendix-proof}
\setcounter{lem}{0}
\begin{lem}
The minimum makespan of the LP relaxation of model~\eqref{ct-ooe} with and without mode-consistency constraints~\eqref{ct-ooe-mode-consistency} is equal to zero.
\end{lem}

\begin{proof}
We prove the first part of the result (without inclusion of~\eqref{ct-ooe-mode-consistency}) by constructing a solution with zero makespan that is feasible in the LP relaxation of model~\eqref{ct-ooe}. To this, let $z_{ime}=\frac{1}{|\sMi|\cardA}$ for all~$i\in\sA,m\in\sMi,e\in\sSetminus{\sE}{\cardA}$, $s_e=0$ for all~$e\in\sE$, and~$r_{ik}=\sum_{m\in\sMi}w_{imk}z_{ime}$. We need to show that the solution~$(r,s,u)$ is consistent with every (in-)equality constraint defined in model~\eqref{ct-ooe}. This is trivial for constraints~\eqref{ct-ooe-first-event-start} and \eqref{ct-ooe-event-order}. 

Satisfaction of~\eqref{ct-ooe-nr-resource} is a direct consequence of the definition of~$r_{ik}$. Moreover, we obtain~\eqref{ct-ooe-nr-resource2} since
\[
\sum_{i\in\sA} r_{ik}=\sum_{i\in\sA}\sum_{m\in\sMi}w_{imk}z_{ime}=\sum_{i\in\sA}\sum_{m\in\sMi} w_{imk}\frac{1}{|\sMi|\cardA} \leq W_k \quad k\in\sN,
\]
where the last inequality is true because otherwise there must be at least one activity-mode combination~$(i,m)$ and non-renewable resource~$k$ such that~$w_{imk}>W_k$, which contradicts the fact that there exists, by assumption, no infeasible activity-mode combination. Constraints~\eqref{ct-ooe-resource} follow in a similar manner.

As to~\eqref{ct-ooe-event-time-def}, it is more convenient to analyze the stronger  inequalities~\eqref{ct-ooe-start-time-strong}. That is, we obtain
\[
p_{im}(z_{imf} - z_{ime}-z_{img}) \leq 0 = s_g-s_e \quad i\in\sA,m\in\sMi;e,f,g\in\sE:e<f<g.
\]
Regarding~\eqref{ct-ooe-atleast-one-event}-\eqref{ct-ooe-precedence}, feasibility follows from similar arguments as in the \ac{sm} case (i.e., if~$|\sMi|=1, i\in\sA$). For example, the covering-type constraints~\eqref{ct-ooe-atleast-one-event} are obtained via
\[
\sum_{m\in\sMi}\sum_{e\in\sSetminus{\sE}{\cardA}}z_{ime} = \sum_{m\in\sMi}\sum_{e\in\sSetminus{\sE}{\cardA}}\frac{1}{|\sMi|\cardA} = 1 \geq 1 \quad i\in\sA.
\]
We skip the remaining constraints, which shows the first part of this lemma. For the second part, we must show that constraints~\eqref{ct-ooe-mode-consistency} are satisfied. Indeed, we have
\begin{align*}
\sum_{m'\in\sSetminus{\sMi}{m}}\sum_{f\in\sSetminus{\sE}{\cardA}} z_{im'f} \leq{}&
\sum_{m'\in\sSetminus{\sMi}{m}}\sum_{f\in\sSetminus{\sE}{\cardA}} z_{im'f} + \sum_{f\in\sE\setminus\{e,\cardA\}} z_{imf}\\
={}& \sum_{m'\in\sMi}\sum_{f\in\sSetminus{\sE}{\cardA}} z_{im'f} - z_{ime}\\
={}&1-z_{ime}\\
\leq{}&\cardA\left(1 - z_{ime}\right) \qquad\qquad i\in\sA,m\in\sMi,e\in\sSetminus{\sE}{\cardA}.
\end{align*}
This completes the proof.
\end{proof}


\section*{Acknowledgement}
This work was supported by the Federal Ministry of Education and Research (BMBF) within the project \enquote{DPNB} under grant number 02P17D066.\\ Four anonymous reviewers are thanked for their suggestions which helped clarify and improve earlier versions of the manuscript.

\biboptions{authoryear}

\end{document}